\documentclass[runningheads]{llncs}
\usepackage[T1]{fontenc}
\usepackage{graphicx, multirow}
\usepackage{hyperref}
\usepackage{cleveref}
\usepackage{subcaption}
\usepackage{array}
\usepackage{cite}

\begin{document}
%
\title{Advancing VAD Systems Based on Multi-Task Learning with Improved Model Structures}
%
\titlerunning{VAD Systems}
%
\author{Lingyun Zuo \and Keyu An \and Shiliang Zhang \and Zhijie Yan}
\authorrunning{L. Y. Zuo et al.}
%
\institute{Speech Lab, Alibaba Group, China
\email{\{ailsa.zly,ankeyu.aky,sly.zsl,zhijie.yzj\}@alibaba-inc.com}}
\maketitle              
\begin{abstract}
In a speech recognition system, voice activity detection (VAD) is a crucial frontend module. Addressing the issues of poor noise robustness in traditional binary VAD systems based on DFSMN, the paper further proposes semantic VAD based on multi-task learning with improved models for real-time and offline systems, to meet specific application requirements. Evaluations on internal datasets show that, compared to the real-time VAD system based on DFSMN, the real-time semantic VAD system based on RWKV achieves relative decreases in CER of 7.0\%, DCF of 26.1\% and relative improvement in NRR of 19.2\%. Similarly, when compared to the offline VAD system based on DFSMN, the offline VAD system based on SAN-M demonstrates relative decreases in CER of 4.4\%, DCF of 18.6\% and relative improvement in NRR of 3.5\%.

\end{abstract}
\section{Introduction}
\label{sec:intro}
Automatic Speech Recognition (ASR) systems are becoming increasingly important in human-machine communication.  It is often paired with a voice activity detection (VAD) system~\cite{speech_non_speech_1,speech_non_speech_2,speech_non_speech_3} , which extracts actual speech data from an input audio signal by removing non-speech data before ASR decoding. This enables the ASR system can focus on processing the valid speech segments, reducing recognition errors caused by invalid speech. 

Early works on VAD relied on hand-crafted acoustic features, such as energy ratio, zero-crossing rate, and signal periodicity~\cite{energy_vad_1,energy_vad_2,zero_rate}. Later, supervised machine-learning methods, including Hidden Markov Models (HMM) and Gaussian Mixture Models (GMM), were also shown to be effective~\cite{GMM}. Currently, VAD models based on neural network, such as deep neural networks (DNN)~\cite{DNN}, feedforward sequential memory networks (FSMN)~\cite{FSMN} and Deep FSMN~\cite{DFSMN}, have gained significant attention in speech research. In our system practice, the binary classification model structure has also evolved from DNN to DFSMN, resulting in significant improvements. As application scenarios become more complex, semantic-based VAD systems~\cite{shi2023semantic} have been proposed to improve performance in noisy environments and reduce the high latency in real-time scenarios.

Based on different model structures, our VAD systems have undergone an evolution from single-task binary classification to multi-task semantic classification to enhance the robustness. The rest of this paper is orgainzed as follows. Section 2 presents the practice of VAD system. Section 3 shows the experimental setup, followed by results in Section 4. Finally, Section 5 concludes this work.

\section{Practice of VAD System}
\subsection{Single-task Binary VAD}
\begin{figure}[ht]
  \centering
  \includegraphics[width=1.0\linewidth]{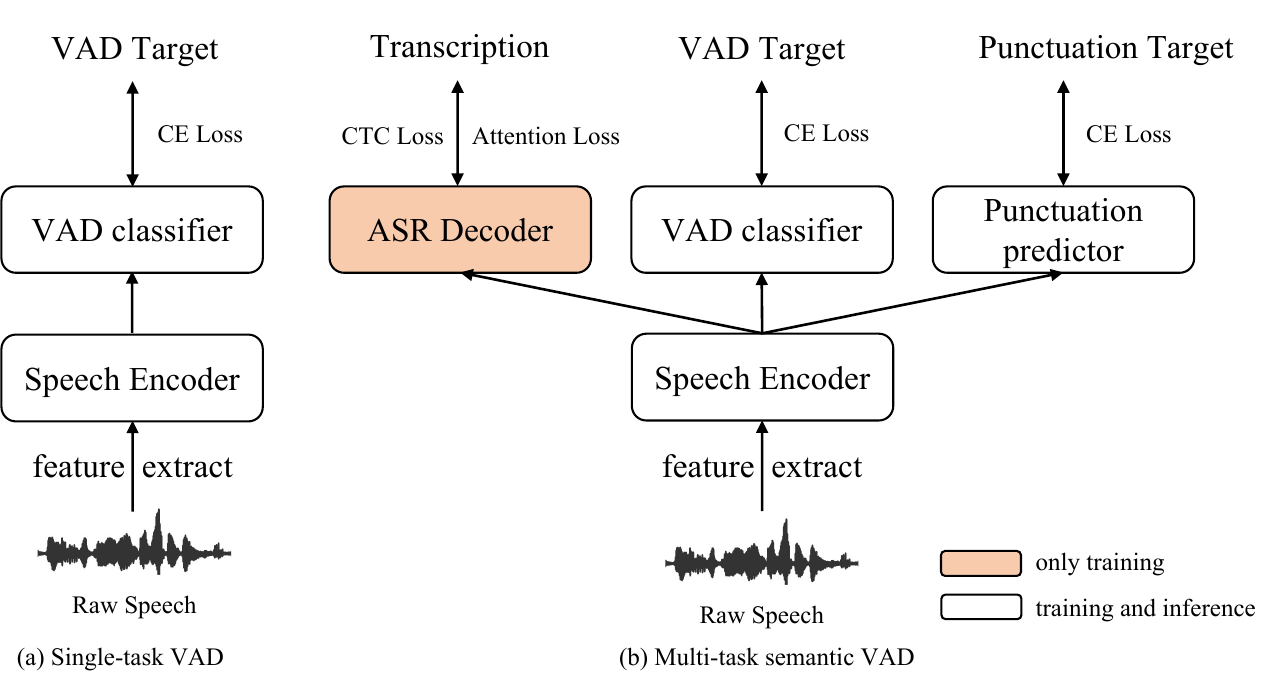}
  \caption{The schematic of VAD training, where the ASR decoder only appears for training in multi-task.}
  \label{fig:vad_training}
  \vspace{-0.25cm}
\end{figure}

The schematic for traditional single-task binary VAD training, as shown in Fig.\ref{fig:vad_training} (a), includes the speech encoder and VAD classifier. Typically, speech encoders employ statistical models like DNN,  FSMN, etc. Furthermore, Cross-Entropy (CE) loss is used for the model training.

\begin{figure}[ht]
  \centering
  \includegraphics[width=0.5\linewidth]{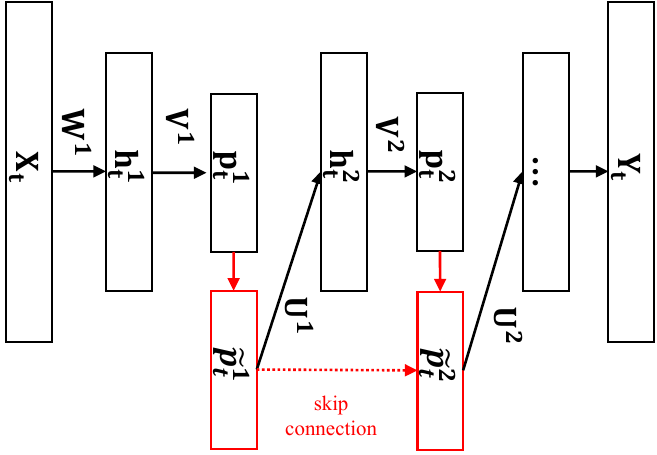}
  \caption{Illustration of Deep-FSMN.}
  \label{fig:dfsmn}
  \vspace{-0.25cm}
\end{figure}

Initially, we employed a DNN-based binary classification model. To improve detection accuracy, we attempt to upgrade the DNN to DFSMN model, which introduces historical information, skip connections between memory units in adjacent hidden layers, addressing the gradient vanishing problem commonly encountered in deep models, as illustrated in Fig.\ref{fig:dfsmn}. The results showed that the DFSMN can achieve relative decreases in Detection Cost Function (DCF) of 34\% with the same size of DNN.

\[
\tilde{\mathbf{p}}^l_t = H(\tilde{\mathbf{p}}^{l-1}_t) + \mathbf{p}_t +
\sum_{i=0}^{N^l_1} \mathbf{a}^l_i \odot \mathbf{p}^l_{t-s_1 \cdot i} +
\sum_{j=1}^{N^l_2} \mathbf{c}^l_j \odot \mathbf{p}^l_{t+s_2 \cdot j}
\]
The formulation of the memory block in DFSMN is shown as above.
Here, $\mathbf{p}^l_t = \mathbf{V}^l\mathbf{h}^l_t + \mathbf{b}^l$
denotes the linear output of the $l$-th linear
projection layer. $\tilde{\mathbf{p}}^l_t$
denotes the output of the $l$-th memory block, ${N^l_1}$ and ${N^l_2}$ denotes the look-back order and lookahead order of $l$-th memory block, respectively~\cite{DFSMN}. 

In this paper, we mainly discuss the model based on DFSMN. In practical applications, the DFSMN model allows for flexible control of model latency by configuring ${N^l_2}$ while maintaining classification performance. Therefore, this model structure has been widely adopted in various online and offline VAD systems. In online scenarios, ${N^l_2}$ in each layer is typically set to a small value, or even 0, to ensure low latency at the model level. In offline scenarios, it can be set to a larger value, to capture future information.

\subsection{Multi-task Semantic VAD}

With the increasing number of speech recognition applications, the traditional binary VAD system's performance in noisy environments and endpoint latency in interactive scenarios have been limited. Through a multi-task training framework, punctuation prediction and ASR tasks are introduced to enhance the learning of semantic information in VAD training~\cite{shi2023semantic}, thereby improving the overall performance of the VAD system.

The schematic of semantic VAD training based on multi-task is shown in Fig.~\ref{fig:vad_training} (b), which consists of the speech encoder, decoder, punctuation prediction and VAD classifier. In the training of semantic VAD, we initially used the same DFSMN structure as speech encoder. However, we draw inspiration from two popular model structures in ASR domain, Receptance Weighted Key Value(RWKV)~\cite{RWKV} and memory equipped self-attention (SAN-M)~\cite{SANM}, to train real-time and offline VAD system, respectively.

\begin{figure}[ht]
  \centering
  \includegraphics[width=1.0\linewidth]{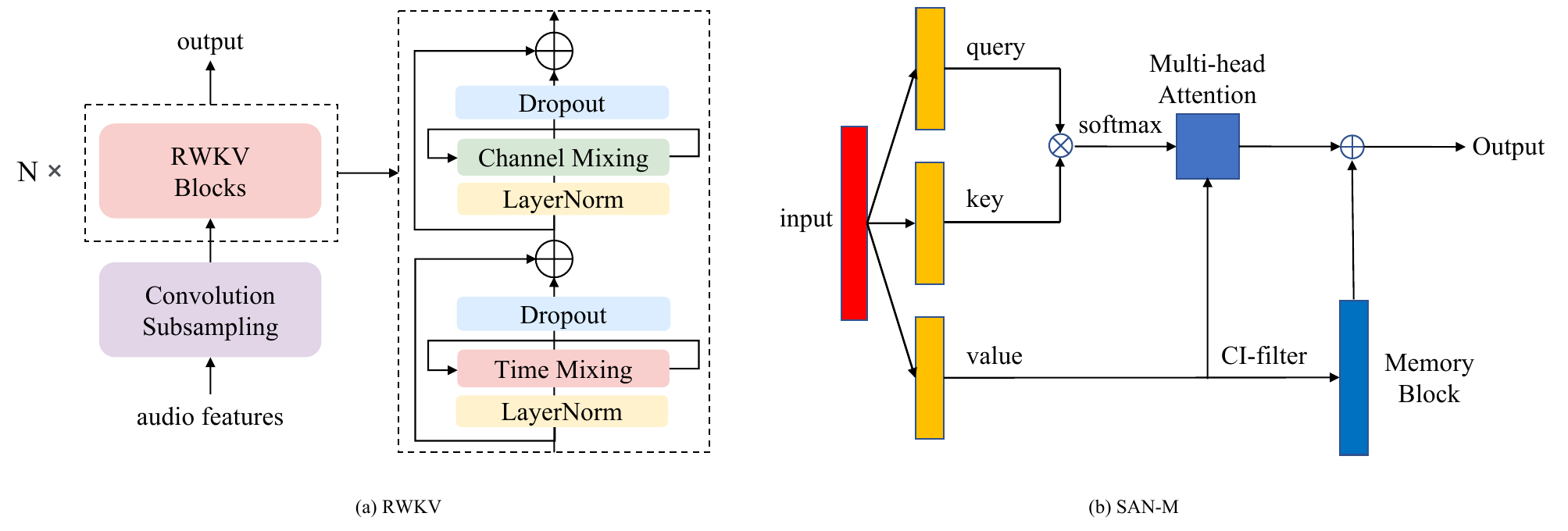}
  \caption{Improved Model architectures used for semantic VAD}
  \label{fig:vad_rwkv_sanm}
  \vspace{-0.25cm}
\end{figure}

The RWKV model architecture combines the advantages of both RNN~\cite{RNN} and Transformer~\cite{transfomer}, it is well-suited for real-time VAD system. We employ an enhanced RWKV model by incorporating a convolution subsampling layer before the RWKV block. Each RWKV block is composed of a time-mixing and a channel-mixing sub-blocks with recurrent structures, as illustrated in Fig.~\ref{fig:vad_rwkv_sanm} (a).

Given the input sequence ${\bf x}$, an RWKV block combines the time mixing module and channel mixing module using: 
$$
{\bf x}^{\prime} = {\bf x} + {\rm Dropout(TimeMixing(LayerNorm({\bf x})))}
$$
$$
{\bf x}^{\prime \prime} = {\bf x}^{\prime} + {\rm Dropout(ChannelMixing(LayerNorm({\bf x^{\prime}})))}
$$
Different from the original formulation, we add a Dropout layer before residual connection to avoid over-fitting.

The SAN-M architecture combines the self-attention ability of Transformer with DFSMN memory block, allowing it to enhance CD-dependencies within a single feature through self-attention and learn CI-dependencies from the statistical average distribution of the entire dataset using DFSMN memory blocks. It has been demonstrated that the SAN-M outperforms Transformer in ASR tasks~\cite{SANM}. The SAN-M architecture is shown in Fig.\ref{fig:vad_rwkv_sanm} (b) that a DFSMN filter has been added on the \textit{values} inside the \textit{Multi-Head Attention} to output.
During the training of SAN-M, since the training data typically has a longer duration ($>$ 10s), it is common practice to chunk the input data, referred to as SAN-M chunks.

\section{Experimental Setup}

\subsection{Dataset}
We conduct extensive experiments to evaluate the performance of DFSMN, RWKV and SAN-M on Mandarin speech recognition task. In the experiments, we use 1500 hours of internal Mandarin speech data as the training set, where each piece of speech has a minimum duration of 10s. We use approximately 10 hours of long speech data as test sets, including general near-field(Test1, Test2, and Test3 in Table~\ref{tab:vad_model_cer_dcf_realtime}), conference data with noise(Test4 and Test5 in Table~\ref{tab:vad_model_cer_dcf_realtime}), and audio-video scenarios(Test6 in Table~\ref{tab:vad_model_cer_dcf_realtime}), etc. 
In addition, we also use approximately 6 hours of pure noise test sets, including about 2000 BGM samples and 4000 general noise samples.

\subsection{Evaluation metrics}
DCF is one of the most commonly used evaluation metrics in VAD systems~\cite{shi2023semantic}. In this paper, we propose a new metric called Noise Rejection Rate (NRR) to evaluate the system's performance based on real-world experience. The NRR is used to assess the system's resistance to noise in pure noise datasets. Furthermore, it is common to assess the impact on CER (Character Error Rate) by combining VAD with ASR on the long speech.

\subsubsection{Noise Rejection Rate(NRR):}
$N$ represents the number of audio samples in the pure noise test set, and $M$ represents the number of these noise audio samples that are rejected by the VAD model. The ratio $\frac{M}{N}$ is referred to as NRR.

\subsection{Model configuration}
In this work, we use the 80-dimensional log-mel filterbank (Fbank) as the input feature. The window size was set to 25ms with a window shift of 10ms, and all frames are downsampling by a factor of 2 before the speech encoder.

The VAD based on DFSMN consists of an input linear layer, ten DFSMN blocks, and an output linear layer. The linear layer has a dimension of 256. The project dimension of DFSMN block is 1024. The lorder is set to 10 and the rorder are set to 0 and 10 for online and offline mode, respectively.

The semantic VAD training consists of four components: encoder, decoder, punctuation prediction and VAD classifier. We compare two improved models, RWKV and SANM, with DFSMN. The configuration of the semantic VAD model based on DFSMN is similar to the binary classification task, with the addition of semantic information. During training, the ASR adopts the CTC loss. In the RWKV, a single Conv2d layer with $out\_channels=256$, $kernel\_size=3$, and $stride=2$ is followed by a linear layer with an output dim set to 256. Following that, there are 4 RWKV blocks. The dimensions of self-attention and the feedforward network in the block are set to 256 and 1024, respectively. The corresponding decoder is a single rnnt layer with hidden size 320.
The SAN-M-based encoder consists of 4 blocks of SAN-M and a feed-forward component with 4-head multi-head attention (MHA). The dimensions of MHA and the feedforward network (FFN) are set to 320 and 1280, respectively. The corresponding decoder is based on Streaming Chunk-Aware Multihead Attention (SCAMA)~\cite{scama}. 

In all experiments, the size of the speech encoder based on DFSMN, RWKV and SAN-M are about 24MB. We use the FunASR toolkit ~\cite{funasr} for our experiments.

\begin{table}[!t]
\centering
\caption{CER(\%) and DCF(\%) of Different VAD Models in Real-time Systems}
\label{tab:vad_model_cer_dcf_realtime}
\setlength{\tabcolsep}{12pt}
\renewcommand{\arraystretch}{1.5}
\vspace{1mm}
\begin{tabular}{c|c|c|c|c|c|c}
\hline
 & \multicolumn{2}{c|}{\textbf{DFSMN}} & \multicolumn{2}{c|}{\textbf{DFSMN-MultiTask}} & \multicolumn{2}{c}{\textbf{RWKV-MultiTask}} \\
 \cline{2-7}
 & CER & DCF & CER & DCF & CER & DCF \\
\hline
Test1 & 2.53 & 5.13 & 2.44 & 4.87 & 2.30 & 4.98 \\
Test2 & 4.31 & 1.50 & 4.47 & 1.71 & 4.10 & 1.55 \\
Test3 & 6.95 & 6.07 & 6.51 & 2.42 & 6.36 & 1.99 \\
Test4 & 19.99 & 10.49 & 18.82 & 7.42 & 18.35 & 6.75 \\
Test5 & 20.60 & 10.06 & 18.90 & 5.57 & 18.47 & 4.37 \\
Test6 & 16.00 & 16.71 & 15.54 & 16.71 & 15.85 & 17.30 \\
\hline
\textbf{AVG} & 11.73 & 8.33 & 11.11 & 6.45 & 10.91 & 6.16 \\
\hline
\end{tabular}
\end{table}

\begin{table}[!t]
\centering
\caption{CER(\%) and DCF(\%) of Different VAD Models in Offline Systems}
\label{tab:vad_model_cer_dcf_offline}
\setlength{\tabcolsep}{12pt}
\renewcommand{\arraystretch}{1.5}
\vspace{1mm}
\begin{tabular}{c|c|c|c|c|c|c}
\hline
 & \multicolumn{2}{c|}{\textbf{DFSMN}} & \multicolumn{2}{c|}{\textbf{DFSMN-MultiTask}} & \multicolumn{2}{c}{\textbf{RWKV-MultiTask}} \\
 \cline{2-7}
 & CER & DCF & CER & DCF & CER & DCF \\
\hline
Test1 & 2.61 & 4.65 & 2.53 & 4.68 & 2.33 & 4.83 \\
Test2 & 4.37 & 0.97 & 4.08 & 1.04 & 4.06 & 1.05 \\
Test3 & 7.09 & 4.84 & 6.71 & 3.31 & 6.42 & 2.37 \\
Test4 & 18.86 & 8.96 & 18.58 & 7.31 & 18.27 & 6.52 \\
Test5 & 20.79 & 9.11 & 19.60 & 6.74 & 19.13 & 5.23 \\
Test6 & 14.66 & 14.89 & 15.16 & 15.93 & 15.16 & 15.31 \\
\hline
\textbf{AVG} & 11.40 & 7.24 & 11.11 & 6.50 & 10.90 & 5.89 \\
\hline
\end{tabular}
\end{table}

\section{Results}
In real-time systems, we present the results of CER, DCF, and NRR in Table~\ref{tab:vad_model_cer_dcf_realtime} and Table~\ref{tab:vad_model_nrr_realtime}. 
Compared to the binary classifier based on DFSMN, the semantic VAD based on RWKV achieves relative decreases in CER of 7.0\%, DCF of 26.1\% and relative improvement in NRR of 19.2\%.

In offline systems, we present the results of CER, DCF, and NRR in Table~\ref{tab:vad_model_cer_dcf_offline} and Table~\ref{tab:vad_model_nrr_offline}. Compared to the binary classifier based on DFSMN, the offline VAD system based on SAN-M demonstrates relative decreases in CER of 4.4\%, DCF of 18.6\% and relative improvement in NRR of 3.5\%.

Overall, semantic models based on DFSMN have shown significant improvements compared to traditional binary classification. Furthermore, the performance of the semantic models can be further enhanced by incorporating improved model structures such as RWKV and SAN-M.
For example, on the Test5 in Table\ref{tab:vad_model_cer_dcf_realtime}, which is a noisy conference dataset, a relative decrease of 10\% in CER is achieved.
However, the CER of Test6 in tabel\ref{tab:vad_model_cer_dcf_offline} has deteriorated due to the lack of contextual data. In the future, optimizations will be made based on specific cases, including training data and post-processing strategies.


\begin{table}[!t]
\centering
\caption{NRR(\%) of Different VAD Models in Real-time Systems}
\label{tab:vad_model_nrr_realtime}
\setlength{\tabcolsep}{12pt}
\renewcommand{\arraystretch}{1.5}
\vspace{1mm}
\begin{tabular}{c|c|c|c}
\hline
     &{\textbf{DFSMN}} & {\textbf{DFSMN-MultiTask}} & {\textbf{RWKV-MultiTask}} \\
\hline
BGM & 85.28 & 92.46 & 95.34 \\
Noise & 29.75 & 35.23 & 41.79 \\
\hline
\textbf{AVG} & 57.51 & 63.84 & 68.56 \\
\hline
\end{tabular}
\end{table}

\begin{table}[!t]
\centering
\caption{NRR(\%) of Different VAD Models in Offline Systems}
\label{tab:vad_model_nrr_offline}
\setlength{\tabcolsep}{12pt}
\renewcommand{\arraystretch}{1.5}
\vspace{1mm}
\begin{tabular}{c|c|c|c}
\hline
& {\textbf{DFSMN}} & {\textbf{DFSMN-MultiTask}} & {\textbf{SANM-MultiTask}} \\
\hline
BGM & 97.22 & 97.70 & 97.85 \\
Noise & 58.24 & 57.90 & 63.05 \\
\hline
\textbf{AVG} & 77.73 & 77.80 & 80.45 \\
\hline
\end{tabular}
\end{table}

\section{Conclusions}
In this paper,
we present the real-time semantic VAD system based on RWKV and the offline semantic VAD system based on SAN-M. Experimental results show that the semantic VAD systems outperforms the DFSMN-based system in terms of CER, DCF, and NRR metrics. In the future, we will explore the utilization of more complex data and employ advanced post-processing strategies to further enhance the performance of the semantic VAD system.

\end{document}